\documentclass[
reprint,
superscriptaddress,
tightenlines,
aps,
prl,
floatfix
]{revtex4-1}

\usepackage{amsmath}
\usepackage{amssymb}
\usepackage{amsthm}
\usepackage{chngcntr}
\usepackage{dcolumn}
\usepackage{float}
\usepackage{graphicx} 
\usepackage[colorlinks=true, allcolors=blue]{hyperref}
\usepackage{lipsum}
\usepackage{tabularray}
\begin{document}

\title{Ringdown analysis of GW250114 with \texttt{RingCWB}}
\author{Achal Kumar}
\email{achal.kumar@ufl.edu}
\affiliation{Department of Physics, University of Florida,
Gainesville, Florida, USA}

\author{Poulami Dutta Roy}
\email{duttaroy.poulami@ufl.edu}
\affiliation{Department of Physics, University of Florida,
Gainesville, Florida, USA}

\author{Sergey Klimenko}
\email{klimenko@phys.ufl.edu}
\affiliation{Department of Physics, University of Florida,
Gainesville, Florida, USA}

\begin{abstract}
The binary black hole merger GW250114, the loudest gravitational-wave signal detected to date, offers an opportunity to perform a precise test of Einstein's general relativity (GR). Its exceptionally high signal-to-noise ratio enables accurate measurements of the quasi-normal modes (QNMs) and properties of the remnant black hole. In this work, we present a ringdown analysis of GW250114 using the \texttt{RingCWB} algorithm. We measure the dominant $(2,2,0)$ QNM along with its first overtone and constrain the fractional deviations in frequency and damping time from their GR-predicted values, finding $\delta f_{220} = 0.004_{-0.023}^{+0.022}$ and $\delta\tau_{220} = 0.010_{-0.087}^{+0.096}$ , both consistent with zero within the quoted uncertainties. These results represent stringent constraints on deviations from GR. The analysis also places tight constraints on the remnant spin, consistent with a binary black hole formed through stellar evolution.

\end{abstract}

\maketitle
\paragraph{Introduction}
On January 14, 2025, The Laser Interferometer Gravitational-Wave Observatory (LIGO)~\cite{LIGOScientific:2014pky} recorded the loudest gravitational-wave (GW) signal to date, GW250114\_082203 (hereafter GW250114)~\cite{LIGOScientific:2025wao,LIGOScientific:2025rid} with a network signal-to-noise ratio (SNR) of $80$.  The high SNR of GW250114 provides a powerful probe to test whether Einstein's theory of general relativity (GR) describes the  observed gravitational radiation. Among the stages of a GW signal from a Binary Black Hole (BBH) merger, the ringdown phase provides one of the cleanest avenues for testing GR. Ringdown occurs at the end of the BBH merger where the perturbed remnant formed as a result of the merger stabilizes into a Kerr black hole (BH) while emitting gravitational waves in the process. Ringdown is governed by the perturbative dynamics of the remnant black hole, which is well described within GR~\cite{Berti:2025hly,Chandrasekhar:1975zza}. A linearly perturbed Kerr BH is expected to emit GW radiation described by quasinormal modes (QNMs), which are damped harmonic signals characterised by specific frequencies and damping times~\cite{Vishveshwara:1970zz,Berti:2009kk}. The QNMs represent the characteristic BH spectrum that uniquely identifies it as described by GR.  A precise measurement of this spectrum would provide a stringent test of GR and a direct insight into the nature of BHs. 

Previous ringdown analyses of GW250114~\cite{LIGOScientific:2025wao,LIGOScientific:2025rid} perform tests of GR using different methods, including \texttt{pyRing}~\cite{Gennari:2023gmx} and \texttt{pSEOBNR}~\cite{Brito:2018rfr}. While \texttt{pyRing}  fits ringdown models to a selected post-merger segment of the strain data, \texttt{pSEOBNR} analyzes the full inspiral–merger–ringdown signal using a parameterized SEOBNRv5PHM waveform model~\cite{Ramos-Buades:2023ehm,Pompili:2023tna} that allows deviations in QNMs frequencies and damping times from their GR predictions.  

In this work we present an independent ringdown analysis of GW250114 using \texttt{RingCWB}~\cite{kumar2026}, an analysis framework based on coherent WaveBurst (cWB), which
is an unmodeled pipeline for the detection and reconstruction of GW signals \cite{Mishra:2024zzs,Klimenko:2005xv,Klimenko:2008fu,Klimenko:2015ypf}. The reconstructed
cWB waveforms are used for the ringdown parameter estimation instead of the raw strain data. In \texttt{RingCWB},
cWB acts as a denoising algorithm that mitigates the
impact of the detector noise on the ringdown reconstruction, which improves the precision of the measurements. \texttt{RingCWB} performs ringdown analysis approximately $2~t_{M_f}$ after the peak time of the signal, where $t_{M_f}=GM_f/c^3$ is the characteristic timescale set by the detector-frame remnant mass. This allows for utilization of the high SNR signal near the peak amplitude, which further improves the precision of the measurements. The estimated ringdown parameters, including the QNM frequencies and damping times, as well as the detector frame remnant mass $(M_f)$  and spin $(a_f)$, are compared with the predictions of GR obtained from the source properties inferred using the detected GW signal.  The source properties are inferred by matching the full inspiral-merger-ringdown (IMR) signal to simulated GR waveforms yielding posterior probability distributions for the source parameters \cite{Cutler:1994ys,LIGOScientific:2019hgc,Veitch:2014wba,Christensen:2022bxb,LIGOScientific:2025yae}.  These posteriors are used to then calculate  the remnant properties that represent the GR predicted values, against which the independent measurement is compared.

\begin{figure*}[t]
    \centering
    \hspace{-0.3cm}
    \includegraphics[width=\textwidth]{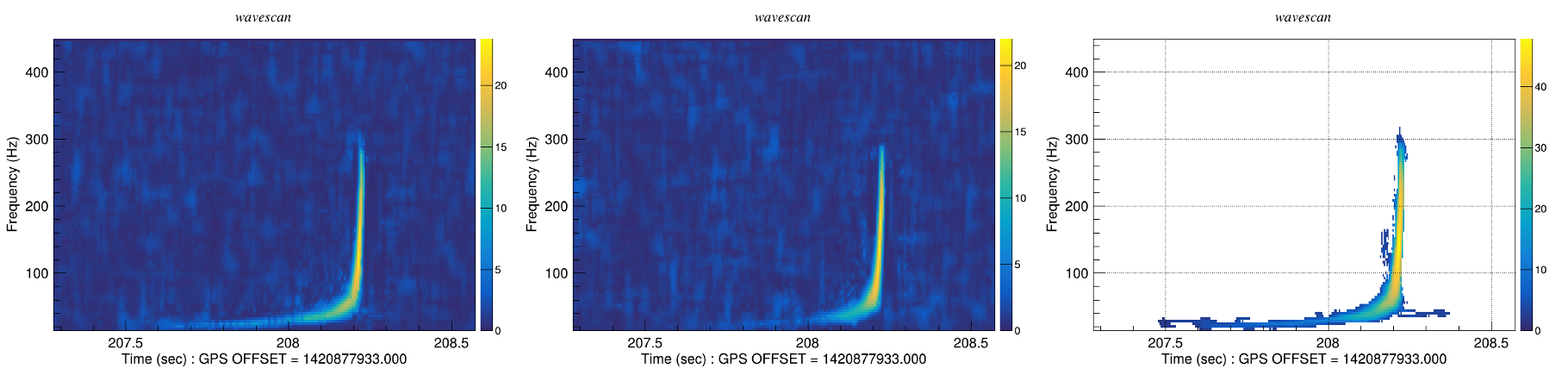}
\caption{\label{fig:ced_250114} Time-frequency representation of the first event GW250114 detected by LIGO: left panel - Livingston detector, middle panel - Hanford detector, right panel - Time-frequency data used to obtain the reconstructed time domain signal with the inverse Wavescan transform~\cite{Klimenko:2022nji}. }
\end{figure*}

\paragraph{Ringdown Model}
The ringdown model is described as a superposition of damped harmonic signal,
\begin{equation}
\label{eq:model}
\begin{split}
h(t) &= Ae^{-t/\tau_{220} + i[2\pi f_{220}t + \phi_{220}]} \\
&\quad + A\sum_{lmn} \epsilon_{lmn}
e^{ -t/\tau_{lmn} +  i[2\pi f_{lmn}t + \phi_{lmn}]} ,
\end{split}
\end{equation}
where $A$ denotes the amplitude of the dominant $(2,2,0)$ mode. The parameters $\epsilon_{lmn}$ represent the relative amplitudes of the subdominant modes, such that $A\epsilon_{lmn}$ gives the amplitude of the  $(l,m,n)$ mode. 
For each mode, $f_{lmn}$, $\tau_{lmn}$, and $\phi_{lmn}$ denote the frequency, damping time, and phase, respectively. 

The analysis is performed using three different approaches. 
First, we consider a ringdown model comprising the dominant $(2,2,0)$ mode and its first overtone $(2,2,1)$~\cite{kumar2026,LIGOScientific:2025wao}.
The model is parameterized by the remnant mass $M_f$ and spin $a_f$, from which the QNM frequencies and damping times are determined using the Kerr spectrum. The test of GR is then performed by evaluating the fractional deviations as defined in Eq.~(10) of \cite{kumar2026}. Second, the ringdown model is extended by including additional QNMs allowing us to search for subdominant modes beyond the $(2,2,1)$ overtone. Third, we consider the two  modes $(2,2,0)$ and $(2,2,1)$ with independent frequencies and damping times enabling an independent measurement of these modes.


\begin{figure}[htbp]
    \centering      \includegraphics[width=0.45\textwidth]{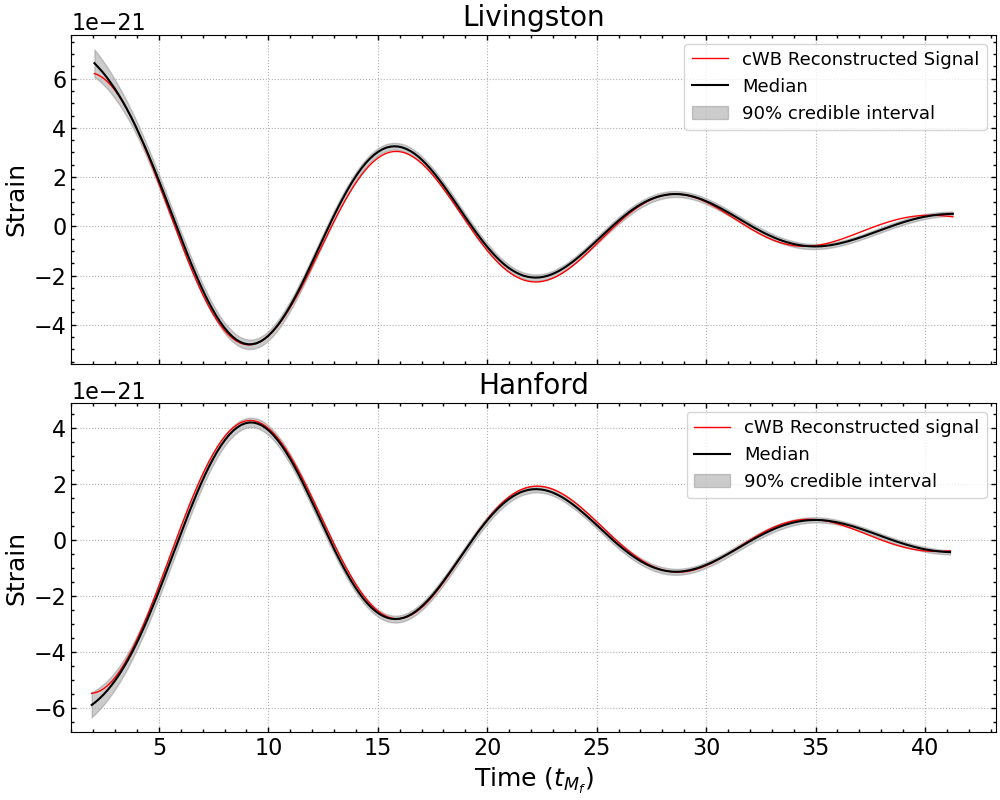} \caption{\label{fig:fit} The GW250114 signal reconstructed with cWB (red)  along with the ringdown model evaluated at the median values (black). The gray regions indicate the $90\%$ credible intervals of the ringdown model. } 
\end{figure}

\paragraph{Results} The first approach provides the most stringent constraints on the ringdown parameters among the three approaches, owing to its smaller parameter space. Fig.~\ref{fig:fit} shows the segment of cWB-reconstructed GW250114 signal used in the ringdown analysis, together with the corresponding best-fit ringdown model, that is  parameterized in terms of the remnant mass and spin. The shaded regions indicate the $90\%$ confidence intervals of the model. The reconstructed signal lies within the $90\%$ confidence band, indicating that the data is well described by the model.

\begin{figure}[htbp]
\hspace{-0.7cm}
 \includegraphics[width=0.5\textwidth]{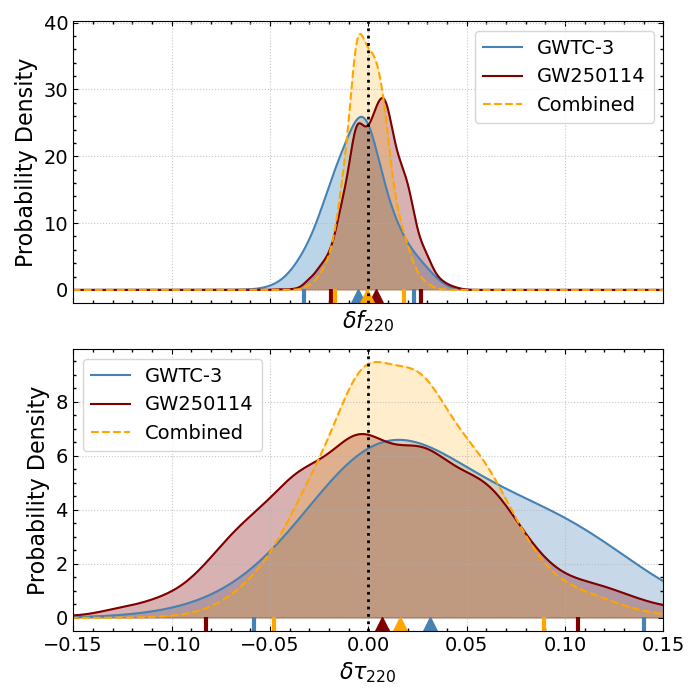} \caption{\label{fig:f_tau} Probability density of $(2,2,0)$ QNM frequency deviation $\delta f_{220}$ (top) and damping time deviation $\delta \tau_{220}$ (bottom), for GW250114 and the combined results obtained using GWTC-3 \cite{kumar2026}. The triangles mark the median values and  the vertical bars mark the $90\%$ credible intervals.
 }
\end{figure}

The measured $(2,2,0)$ mode frequency $f_{220}$ and damping time $\tau_{220}$ are compared with the corresponding IMR parameters,  to calculate the fractional deviations from GR as defined in Eq.~(10) of \cite{kumar2026}. The resulting deviations are 
\begin{equation}
\label{eq:deviations}
\delta f_{220} =0.004_{-0.023}^{+0.022}, \\ \quad 
\delta \tau_{220} = 0.010_{-0.087}^{+0.096},
\end{equation}
where the quoted uncertainties correspond to $90\%$ confidence intervals. The high SNR of GW250114 leads to stringent bounds on the deviations which are consistent with zero. 
The accuracy achieved with this single event is comparable to that obtained in our previous analysis of the GWTC-3 events~\cite{kumar2026}.  These deviations in $(2,2,0)$ QNM are consistent  with those obtained by the \texttt{pSEOBNR} analysis~\cite{LIGOScientific:2025wao}, with uncertainties of comparable magnitude. Figure~\ref{fig:f_tau} shows the posterior distributions of the deviation parameters for GW250114 and the corresponding joint posteriors from the GWTC-3 analysis. 
Combining GW250114 with the GWTC-3 events further improves the constraints on the deviation parameters, yielding the following bounds, 
\begin{equation}
    \delta f_{220} =-0.001_{-0.016}^{+0.018} \quad  \delta \tau_{220} = 0.016_{-0.064}^{+0.073}.
\end{equation}

Table~\ref{tab:ring} shows the remnant properties estimated by the ringdown analysis. The measured value of the relative amplitude  $\epsilon_{221}$ deviates significantly from zero, showing that the $(2,2,1)$ QNM has a significant contribution to the ringdown signal. The analysis also provides a stringent constraint on the remnant spin which is consistent with the canonical value of $a_f \sim 0.68$ expected for black holes formed through stellar evolution.

\begin{table}[htbp]
\centering
\renewcommand{\arraystretch}{1.3} 
\setlength{\tabcolsep}{10pt}
\begin{tabular}{l r}
\hline
\hline
Remnant Mass~$(M_f/M_{\odot})$  & $68.2_{-3.7}^{+3.9}$  \\
Remnant Spin~$(a_f)$  & $0.68_{-0.09}^{+0.07}$  \\
$f_{220}$~(Hz)             & $249.0_{-5.7}^{+5.4}$ \\
$\tau_{220}$~(ms)         & $4.1_{-0.3}^{+0.4}$  \\
$\epsilon_{221}$            & $0.75_{-0.11}^{+0.11}$ \\
\hline
\end{tabular}

\caption{Ringdown parameters in the detector frame, including the $(2,2,0)$ mode frequency, $f_{220}$, damping time, $\tau_{220}$, detector frame remnant mass, $M_f$, remnant spin, $\chi_f$, and the relative amplitude, $\epsilon_{221}$ of the $(2,2,1)$ QNM.  The quoted uncertainties correspond to $90\%$ credible intervals and include all identified systematic uncertainties~\cite{kumar2026}.}
\label{tab:ring}
\end{table}

\begin{figure}[h]
    \centering      
    \hspace{-0.8cm}    \includegraphics[width=0.52\textwidth]{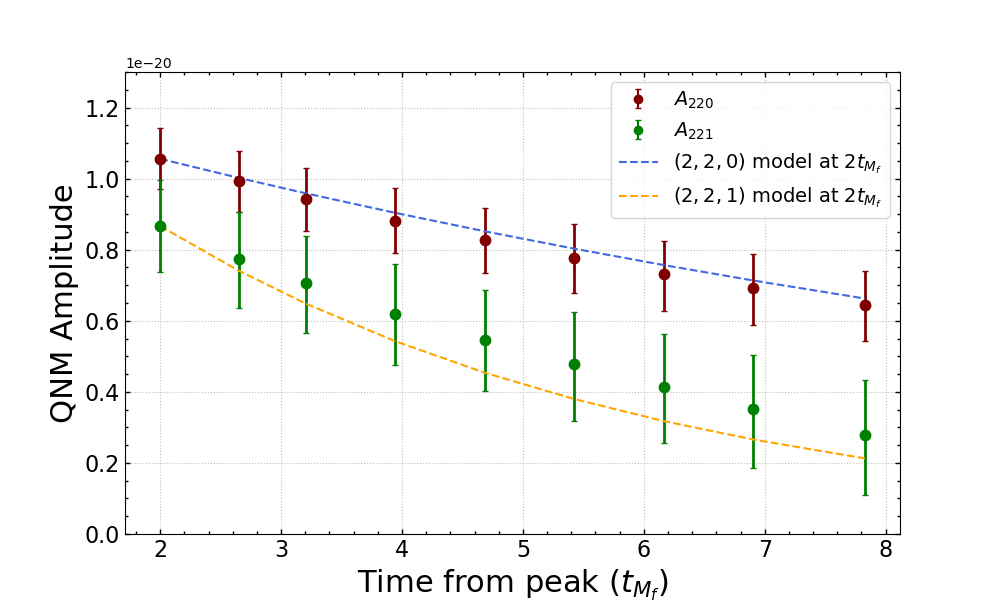} \caption{\label{fig:QNM_amp} The measured  $(2,2,0)$ and $(2,2,1)$ amplitudes as a function of  analysis start time. The curves show the expected exponential decay of the mode amplitudes, calculated using the median ringdown parameters obtained from the standard analysis with a start time of $2~t_{M_f}$.   }
\end{figure}

The \texttt{RingCWB} starts the analysis at $\sim 2~t_{M_f}$ from the time of the peak signal amplitude. To assess the robustness of the results, the analysis is repeated for multiple start times spanning $(2-8)~t_{M_f}$ from the peak time. Fig~\ref{fig:QNM_amp} shows the measured $(2,2,0)$ and $(2,2,1)$ amplitudes as a function of the analysis start time. These measurements are compared with the median ringdown model from the standard analysis starting at $2~t_{M_f}$. The observed variation of the measured mode amplitudes are consistent with the 
expected exponential decay of the ringdown model within the measurement uncertainties. 

Fig~\ref{fig:mult_win} shows the deviation parameters and remnant spin as a function of the analysis start time. 
\begin{figure*}[t]
    \centering   \includegraphics[width=\textwidth]{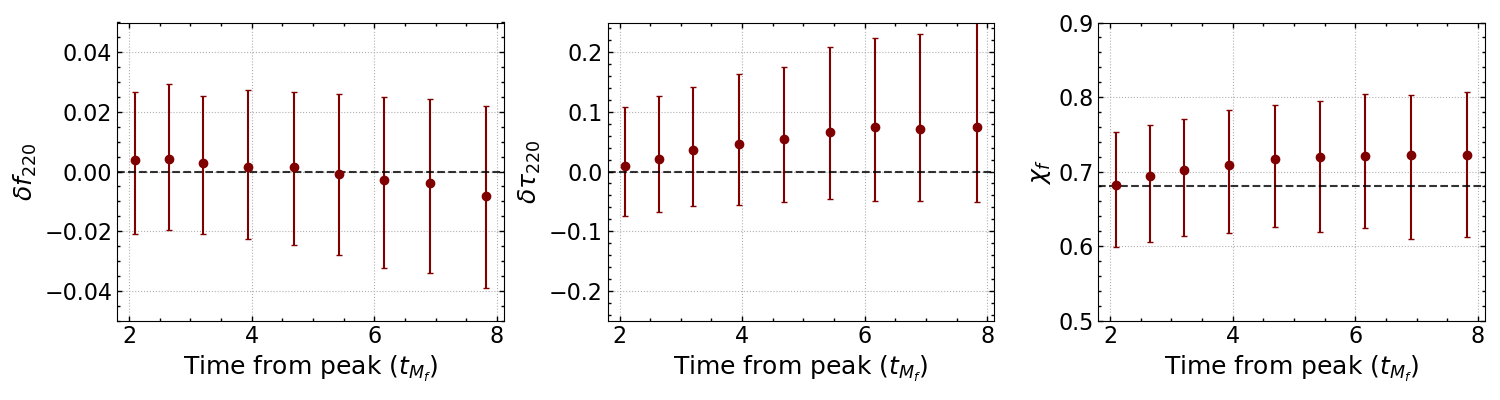}
\caption{\label{fig:mult_win}     The variation of the deviation parameters and remnant spin for different analysis start times relative to the peak time in units of $t_{M_f}$.  }
\label{fig:mult_win}
\end{figure*}
The frequency and damping-time deviations remain consistent with zero, within the measured uncertainties, across  start times of the analysis. Similarly, the remnant spin remains consistent with the canonical value of $a_f \sim 0.68$, across all start times.


\paragraph{Sub-dominant modes}


The second approach described earlier is used to investigate whether the reconstructed signal shows the evidence for subdominant QNMs beyond the $(2,2,1)$ overtone. For GW250114, the QNMs such as the $(3,3)$ and $(4,4)$, which have frequencies above $300$ Hz, are  filtered out by the cWB algorithm, as evident from the right panel of Fig.~\ref{fig:ced_250114}. However modes, including the $(2,2,2)$, $(2,1,0)$, and $(2,0,0)$ could be reconstructed by cWB. We therefore consider three variants of the model, each including one of these modes in addition to the $(2,2,0)$ and $(2,2,1)$ modes. The inferred relative amplitude $\epsilon_{lmn}$ is used to test the presence of these additional QNMs.  The relative amplitudes were found to be $\epsilon_{222} = 0.14^{+0.50}_{-0.41}$, $\epsilon_{210} = -0.09^{+0.57}_{-0.20}$ and $\epsilon_{200} = 0.03^{+0.12}_{-0.13}$. We further compare each extended model with the two-mode $(2,2,0)+(2,2,1)$ model using the log Bayes factor, obtaining $\ln\mathcal{B}^{222} = -0.25$, $\ln\mathcal{B}^{210} = -0.20$, and $\ln\mathcal{B}^{200} = -1.5$ respectively.
These values provide no evidence in favor of any of the additional modes 
beyond the $(2,2,1)$ overtone.

\begin{figure}[h]
    \centering      
    \hspace{-0.7cm} 
    \includegraphics[width=0.45\textwidth]{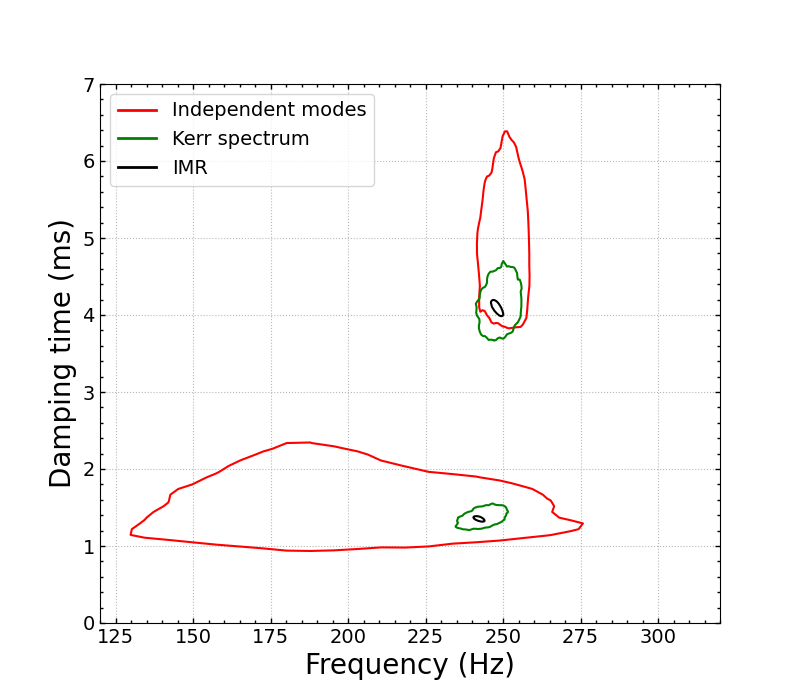} \caption{\label{fig:damp_vs_freq} The $90\%$  credible regions in the damping time-frequency space for the model with two independent QNMs (red) and the  model parametrised using remnant mass and spin using the Kerr QNM spectrum (green), along with the corresponding predictions from the IMR analysis. The contours at the larger damping time correspond to the $(2,2,0)$ mode while the contours at the smaller damping time corresponds to the $(2,2,1)$ mode.  }
\end{figure}

\paragraph{BH spectroscopy}
Even though sub-dominant modes are not observed for GW250114, the fundamental mode $(2,2,0)$  and the first overtone $(2,2,1)$ can be used to perform a test of GR without using the results of the IMR analysis. In this approach, the two QNMs are treated as independent components of the ringdown model, with their frequencies and damping times allowed to vary freely. Thus, no assumption of the Kerr QNM spectrum is made when identifying the modes. Figure~\ref{fig:damp_vs_freq} shows the $90\%$ credible regions in the frequency and damping-time parameter space for the two model choices, together with the corresponding predictions from the IMR analysis. 
The dominant $(2,2,0)$  QNM inferred from this analysis is used to determine the remnant mass and spin, which are then used to predict the frequency and damping time of the $(2,2,1)$ QNM assuming the Kerr spectrum. These predictions are compared with the independently measured $(2,2,1)$ mode to test the consistency of the observed overtone with the Kerr prediction. The discrepancy is quantified  through the fractional deviations $\hat{\delta} f_{221}$  and $\hat{\delta} \tau_{221}$ , which are found to be
\begin{equation}
\hat{\delta} f_{221} = -0.20_{-0.17}^{+0.23},
\qquad
\hat{\delta} \tau_{221} = -0.06_{-0.28}^{+0.28}.
\end{equation}
The deviation parameters show consistency with zero, within the quoted uncertainties. This method yields significantly larger uncertainties because the model with two independent QNMs is considerably less constrained than the model parametrized by the remnant mass and spin. In addition, the $(2,2,1)$ overtone is short-lived, resulting in less precise measurements of its frequency and damping time. Nevertheless, this approach provides a direct measurement of the QNM spectrum and does not rely on the IMR analysis to quantify deviations from the GR. 
\paragraph{Conclusion}
We presented the results of the ringdown analysis of GW250114 using \texttt{RingCWB}. The exceptionally high SNR of this event enabled precise measurements of the ringdown parameters. The measured frequency and damping time of the dominant $(2,2,0)$ mode and its first overtone $(2,2,1)$ showed consistency with the predictions of GR. Moreover, the stringent constraints on the remnant spin of the GW250114 is obtained that show consistency with the predictions of stellar evolution. Finally, we searched for subdominant modes beyond the $(2,2,1)$ mode and found no evidence for their presence.   

\paragraph{Acknowledgment}
This material is based upon work supported by NSF's LIGO Laboratory, which is a major facility fully funded by the National Science Foundation. We gratefully acknowledge the support of LIGO and Virgo for the provision of computational resources, especially LIGO Laboratory, which is supported by the NSF Grant No. PHY $0757058$ and PHY $0823459$. This work was supported by the NSF grant No. PHY-2409372. This research has made use of data, software, and/or web tools obtained from the Gravitational Wave Open Science Center, a service of LIGO Laboratory, the LIGO Scientific Collaboration, and the Virgo Collaboration. We acknowledge the valuable feedback from Tanmaya Mishra, Marek~J.~Szczepa\'nczyk and Gayathri.V.

\bibliography{reference}

\end{document}